\documentclass[trackchanges,twocolumn]{aastex701}

\usepackage{amsmath}	% Advanced maths commands

\begin{document}

\title{Astrometric Calibration of the 4-m International Liquid Mirror Telescope Observations}

\author[0000-0003-1183-9293]{Vibhore Negi}
\altaffiliation{Corresponding author}
\affiliation{Kavli Institute for Astronomy and Astrophysics, Peking University, Beijing, 100871, China}
\affiliation{Aryabhatta Research Institute of Observational Sciences (ARIES), Manora Peak, Nainital, 263002, India}
\email{vibhore.negi18@gmail.com}

\author[0000-0002-0394-6745]{Naveen Dukiya}
\affiliation{Aryabhatta Research Institute of Observational Sciences (ARIES), Manora Peak, Nainital, 263002, India}
\affiliation{Department of Applied Physics, Mahatma Jyotiba Phule Rohilkhand University, Bareilly, 243006, India}
\email{ndukiya@aries.res.in}

\author[]{Gokul Singh Mehra}
\affiliation{Aryabhatta Research Institute of Observational Sciences (ARIES), Manora Peak, Nainital, 263002, India}
\email{gokulmehra51@gmail.com}

\author[0009-0000-1020-9711]{Bhavya Ailawadhi}
\affiliation{Physical Research Laboratory, Navrangpura, Ahmedabad, 380009, India}
\email{bhavya.jk1@gmail.com}

\author[0009-0002-2621-6611]{Monalisa Dubey}
\affiliation{Aryabhatta Research Institute of Observational Sciences (ARIES), Manora Peak, Nainital, 263002, India}
\affiliation{Department of Applied Physics, Mahatma Jyotiba Phule Rohilkhand University, Bareilly, 243006, India}
\email{monalisa@aries.res.in}

\author[0009-0009-3108-3789]{Sara Filali}
\affiliation{Institute of Astrophysics and Geophysics, University of Li{\`e}ge, All{\'ee} du 6 Ao{\^u}t, B-4000 Li{\`e}ge, Belgium}
\email{sfilali@doct.uliege.be}

\author[]{Paul Hickson}
\affiliation{Department of Physics and Astronomy, The University of British Columbia, 6224 Agricultural Road, Vancouver BC V6T 1Z1, Canada}
\affiliation{Outer Space Institute, The University of British Columbia, 6224 Agricultural Road, Vancouver, BC V6T 1Z1, Canada}
\email{hickson@physics.ubc.ca}

\author[]{Brajesh Kumar}
\affiliation{South-Western Institute for Astronomy Research, Yunnan University, Kunming, 650500 Yunnan, China}
\email{brajesharies@gmail.com}

\author[]{Priyanshi Kumari}
\affiliation{Aryabhatta Research Institute of Observational Sciences (ARIES), Manora Peak, Nainital, 263002, India}
\email{kumaripriyanshi98@gmail.com}

\author[0000-0002-4157-5164]{Sapna Mishra}
\affiliation{Space Telescope Science Institute, 3700 San Martin Drive, Baltimore, MD 21218, USA}
\email{sapna.intell@gmail.com}

\author[0000-0003-1637-267X]{Kuntal Misra}
\affiliation{Aryabhatta Research Institute of Observational Sciences (ARIES), Manora Peak, Nainital, 263002, India}
\email{kuntal@aries.res.in}

\author[0000-0001-7783-3797]{Bikram Pradhan}
\affiliation{Indian Space Research Organization, Bengaluru, Karnataka, 560094, India}
\email{bikram8577@gmail.com}

\author[0009-0005-3844-3426]{Kumar Pranshu}
\affiliation{Aryabhatta Research Institute of Observational Sciences (ARIES), Manora Peak, Nainital, 263002, India}
\affiliation{Department of Applied Optics and Photonics, University of Calcutta, Kolkata, 700106, India}
\email{pranshu@aries.res.in}

\author[0000-0002-7005-1976]{Jean Surdej}
\affiliation{Aryabhatta Research Institute of Observational Sciences (ARIES), Manora Peak, Nainital, 263002, India}
\affiliation{Institute of Astrophysics and Geophysics, University of Li{\`e}ge, All{\'ee} du 6 Ao{\^u}t, B-4000 Li{\`e}ge, Belgium}
\email{jsurdej@uliege.be}

\author[0009-0002-2355-5626]{Sarvesh Kumar Yadav}
\affiliation{Aryabhatta Research Institute of Observational Sciences (ARIES), Manora Peak, Nainital, 263002, India}
\affiliation{Department of Applied Optics and Photonics, University of Calcutta, Kolkata, 700106, India}
\email{sarveshkyadav.5201@gmail.com}

% \collaboration[all]{(ILMT Collaboration)} 
%% which tells the compiler how many of the authors above the command to show. For example 

% will include
\begin{abstract}

The 4-m International Liquid Mirror Telescope (ILMT) is a dedicated time domain survey telescope that continuously scans the zenithal sky over the Indian Himalayas in the {\it g$^\prime$, r$^\prime$} and {\it i$^\prime$} optical bands. Its unique capability to repeatedly image the same strip of sky every night makes it a highly useful instrument for the photometric and astrometric studies of Solar System, Galactic and extragalactic objects. We present a robust astrometric calibration pipeline developed for the ILMT data obtained in the time delay integration (TDI) mode. The pipeline uses a linear transformation model from pixel to world coordinates, with a second order correction for the asymmetric optical distortions introduced by the telescope’s optical corrector, and ties the astrometric solution to the Gaia DR3 reference frame. The pipeline is integrated to the routine ILMT data processing workflow. Using data from the first four observing cycles (2022-2025), we present the first assessment of the astrometric performance of the pipeline based on positional residuals of sources cross-matched with Gaia DR3. The pipeline achieves a typical astrometric precision of $\sim$100 milliarcseconds (mas), reaching $\sim$70–80 mas for moderately bright sources (G$\sim$16.5--18.5). These results, based on 347 nights of data, demonstrate the stability and reliability of ILMT astrometry over multi-year timescales. The astrometrically calibrated data from these four observing cycles have been made publicly available to the astronomical community. This work establishes a validated framework for precision astrometry with zenith-pointing TDI surveys and provides a foundation for future time-domain studies with ILMT, including variability characterization, transient localization, and long-term positional monitoring.

\end{abstract}

%% PASP uses Unified Astronomy Thesaurus (UAT) concepts:
% \keywords{\uat{Galaxies}{573} --- \uat{Cosmology}{343} --- \uat{High Energy astrophysics}{739} --- \uat{Interstellar medium}{847} --- \uat{Stellar astronomy}{1583} --- \uat{Solar physics}{1476}}

% \keywords{\uat{Classical Novae}{251} --- \uat{Ultraviolet astronomy}{1736} --- \uat{History of astronomy}{1868} --- \uat{Interdisciplinary astronomy}{804}}

\section{Introduction} 
\label{sec:Intro}
The International Liquid Mirror Telescope (ILMT; \citealt{2025A&A...694A..80S}) is a 4-m zenith-pointing telescope designed for time-domain surveys, located at the ARIES Devasthal Observatory in India. 
The telescope is equipped with a 4k $\times$ 4k CCD camera, with a pixel scale of $\sim0.33
^{\prime\prime}$ pixel$^{-1}$, providing a field of view of $\sim$22.3$^\prime$ $\times$ 22.3$^\prime$ \citep{2022JAI....1140001D}.
It continuously scans a narrow strip of sky centered at the observatory latitude ($+$29$^{\circ}$21$^\prime$41.4$^{\prime\prime}$) in three optical bands ( {\it g$^\prime$}, {\it r$^\prime$}, and {\it i$^\prime$}, \citealt{2022JApA...43...10K,2024BSRSL..93.1054S}),
covering a total sky area of $\sim$120 sq. degrees, and revisiting the same region of sky every night, which makes it well suited to time-domain studies. Observations at zenith offer several advantages, including low observational airmass, optimal image quality, minimal light pollution, and no time lost to pointing adjustments.  
However, as a fixed-pointing liquid mirror telescope, ILMT cannot track the sources transiting across the sky and therefore operates in the time delay integration mode (TDI, also known as drift scanning mode, \citealt{1980SPIE..264...20M,1984MNRAS.210..979H}) for observations \citep{2018BSRSL..87...68S}. 
A dedicated TDI corrector compensates for the star-trail curvature and the differential drift speeds across the field, converting the hyperbolic stellar trajectories into straight lines \citep{1998PASP..110.1081H,2002A&A...388..712V,2024BSRSL..93..863N}. \par
The ILMT reaches limiting magnitudes of $\sim$21.9, 21.7, and 21.5 in {\it g$^\prime$, r$^\prime$,} and {\it i$^\prime$} bands, respectively, as estimated from the initial observations \citep{2024BSRSL..93..820A}. However, under optimal seeing conditions, and with image co-addition, fainter magnitudes are detected with the ILMT. Within these magnitude limits, ILMT is expected to generate a comprehensive database of all the Galactic and extragalactic objects detected within its declination range, providing both astrometric and photometric measurements.

Astrometric calibration of TDI data presents specific challenges owing to the long and narrow image geometry, the continuous drift of sources across the detector, and the presence of scan-dependent distortions. As the scientific return of such surveys depends critically on the accuracy of the astrometric measurements, we have developed a robust pipeline for the astrometric calibration of the TDI mode data obtained with the ILMT. The pipeline is now routinely applied to ILMT observations, and the astrometrically calibrated data from observing cycles $1-4$ (2022-2025) have been made available to the astronomical community \citep{2024BSRSL..93..872M}. \par 
In this paper, we describe the astrometric calibration pipeline for the ILMT data. We start with a brief overview of the ILMT data processing in Section~\ref{sec:data_processing_preview}. In Section~\ref{sec:data_used} we describe the data and catalogs used for developing and testing the pipeline, followed by the methodology in Section~\ref{sec:methodology}.
Section~\ref{sec:results} presents the astrometric performance over the four observing cycles and also examines its dependence on seeing conditions and source brightness, and compares it with results obtained using other approaches. Finally, we summarize our results and conclude in Section~\ref{sec:discussion}. \par
%%%%%%%%%%%%%%%%%%%%%%%%%%%%%
    \begin{figure*}
   \centering
   \includegraphics[width=0.9\linewidth, trim=0 0 0 0, clip]{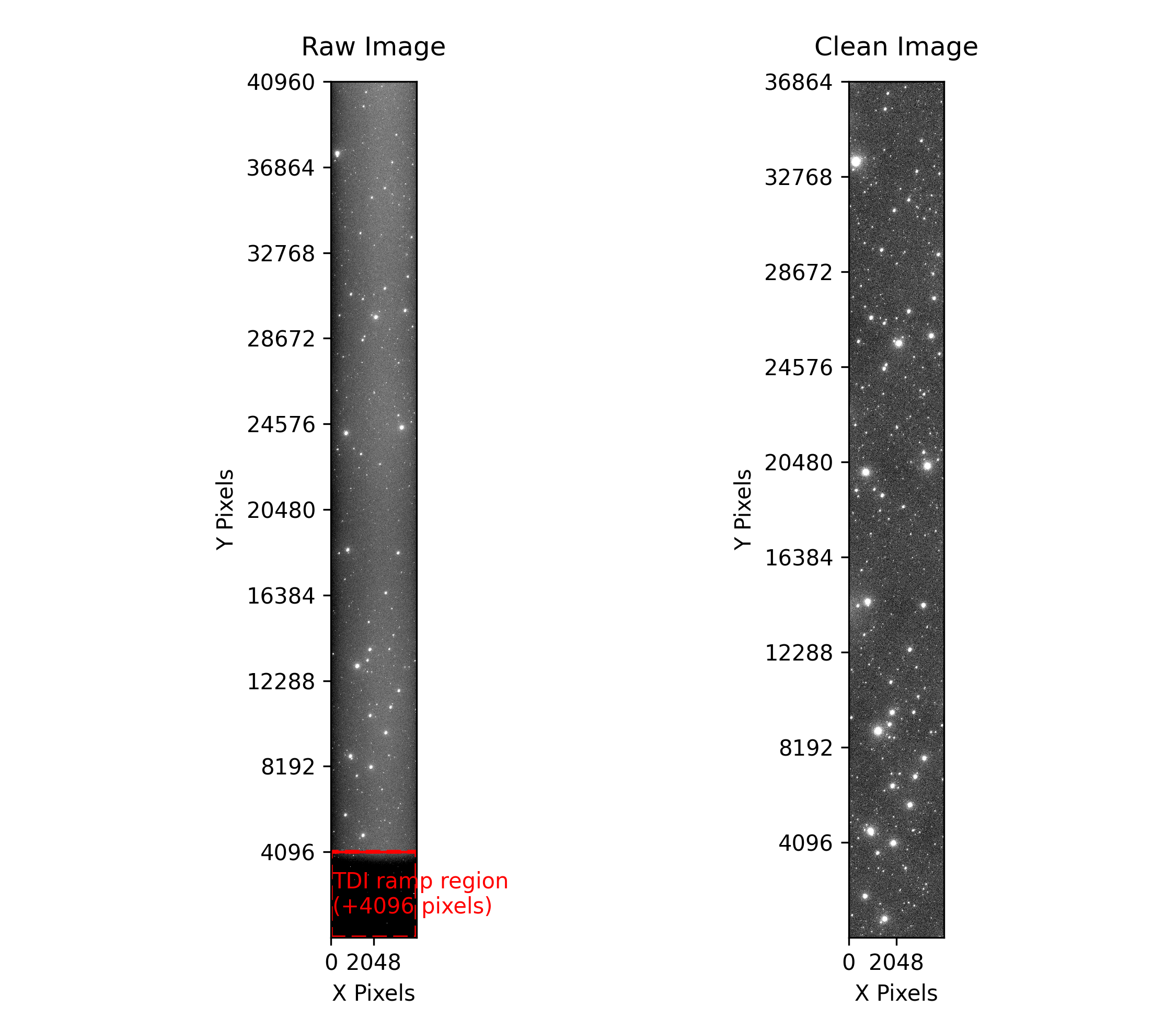}
      \caption{An example raw (left) and clean (right) TDI image obtained with the ILMT on 2025-02-02. The first 4096 pixels along the RA direction (Y pixels) in the raw image, are not fully exposed due to TDI ramping, and hence are removed while cleaning.}
         \label{fig:tdi_ramping}
   \end{figure*}  
%%%%%%%%%%%%%%%%%%%%%%%%%%%%%%
%__________________________________________________________________

    \section{ILMT data processing preview}
\label{sec:data_processing_preview}
The ILMT generates $\sim$15 GBs of data each night \citep{2024BSRSL..93..872M}.
The large amount of the ILMT data requires automatic processing to acquire the raw data, perform their pre-processing, and astrometrically and photometrically calibrate the detected objects, and further perform image subtraction on the science images to detect interesting transient events. The end to end processing of the ILMT data is discussed in \citep{2025A&A...694A..80S}. In brief, each TDI image from the ILMT goes through the following steps, serially:
\begin{enumerate}
\item The raw data are acquired on the {\texttt ccc1} server and transferred to the {\texttt ic1} computer, where the Observatory Control System (OCS) software \citep{2019LPICo2109.6066H} performs real-time pre-processing of each TDI image, applying dark and flat-field corrections, and also discarding the first 22$^{\prime}$ of data along RA which are not fully exposed (see Figure~\ref{fig:tdi_ramping}, also \citealt{2025MNRAS.538..133P} for details). 
\item The pre-processed TDI images are further transferred to the {\texttt icc1} server for further analysis that includes precise astrometric calibration and photometry on the images, and image subtraction. The astrometric calibration pipeline is discussed in detail in Section~\ref{sec:methodology}.
\item The astrometrically calibrated images are further processed for the photometric calibration using a dedicated photometric pipeline for the ILMT (\citealt{2024BSRSL..93..820A}, Ailawadhi et al., in preparation). This pipeline performs both aperture and psf photometry on the ILMT images, and uses the Pan-STARRS1 \citep{2016arXiv161205560C} reference star catalog for the calibration.
\item Finally, the PyLMT pipeline uses these science images and subtracts the reference images to search for any transient events in the images. Any detected new transient candidate is reported to the Transient Name server and may be followed up with the 3.6m Devasthal Optical Telescope, located adjacent to the ILMT \citep{2025MNRAS.538..133P}. \par
The astrometrically calibrated ILMT data is provided to the community at intervals of every six months via the \href{https://www.myqnapcloud.com/share/714efl5j416o5338t7yv1zzz_4d1gf282m0p226q5rsu39664c4cf21if#!/home}{ARIESCloud service} \citep{2024BSRSL..93..872M}.
\end{enumerate}
%%%%%%%%%%%%%%%%%%%%%%%%%%%%%
    \begin{figure*}
   \centering
   \includegraphics[width=0.9\linewidth, trim=0 0 0 0cm, clip]{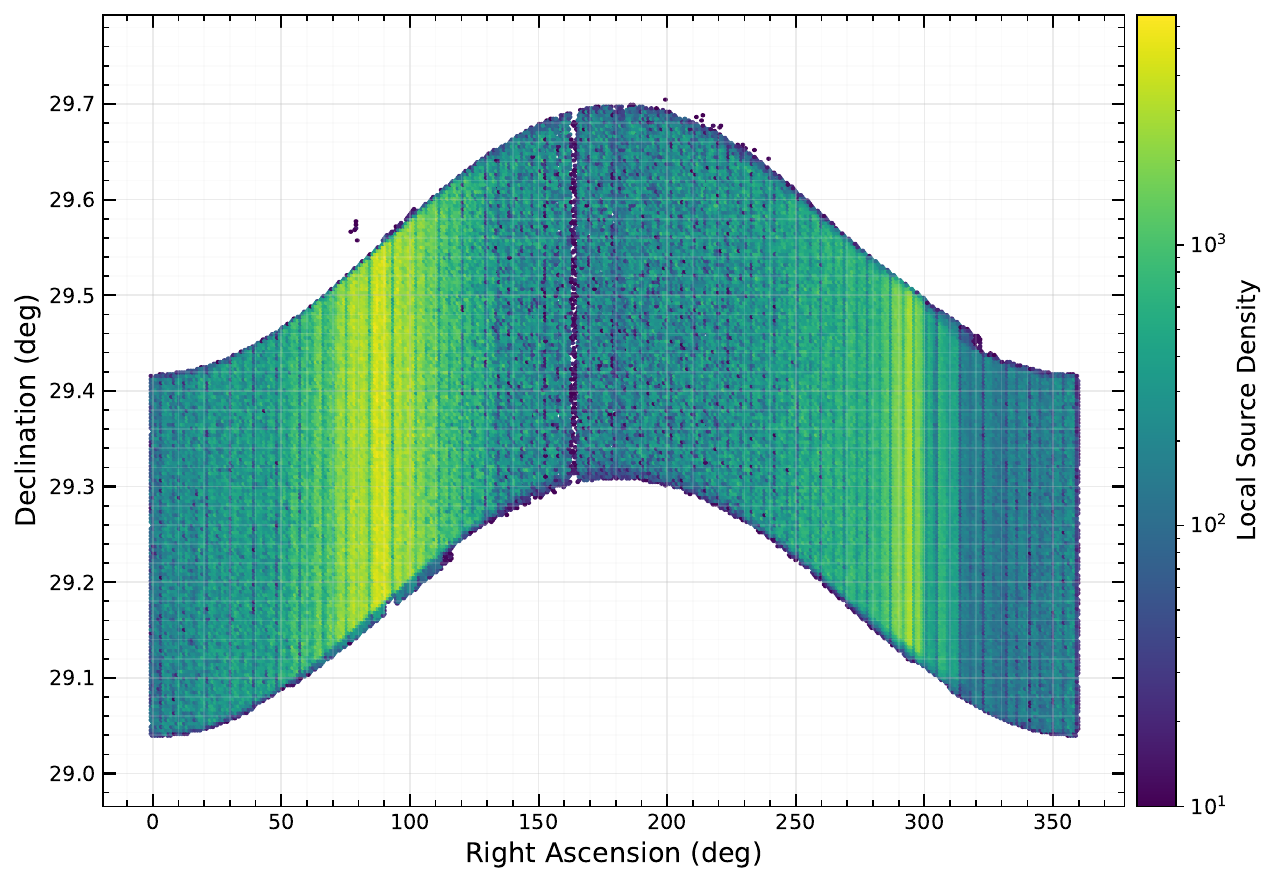}
      \caption{Sky distribution (in J2000) of all the ILMT detections in cycles 1-4. The apparent sinusoidal footprint is a consequence of the observation epoch to J2000 precession transformation, while the strip width is determined by the 22.3$^\prime$ ILMT field of view. The map is generated using hexagonal bins with a gridsize of 360, and the color scale shows log of the number of sources per bin. High-density vertical structures trace the most covered ILMT fields, while low-density regions correspond to fields with lesser nightly coverage. A few scattered detections outside the main band correspond to occasional outliers.}
         \label{fig:result_radec}
   \end{figure*}  
%%%%%%%%%%%%%%%%%
%%%%%%%%%%%%%%%%%%%%%%%%%%%%%%%%%
%%%%%%%%%%%%%%%%%%%%%%%%%%%%%%%%
%%%%%%%%%%%%%%%%%%%%%%%%%%%%%%%%
\section{Data and Catalogs Used}
\label{sec:data_used}
The ILMT started operations in full mode from March 2023, and has completed four observing cycles (April - Nov 2022: Commissioning phase: Cycle 1, March - June 2023: Cycle 2, Nov 2023 - May 2024: Cycle 3, and Nov 2024 - May 2025: Cycle 4) at the time of the writing of this paper. These cycles have produced $\sim$347 nights of useful scientific data and covered a sky area of 115 sq. deg (see Figure~\ref{fig:result_radec} for reference).  
The individual bands contribute to 106, 130, and 111 of observing nights in {\it g$^\prime$}, {\it r$^\prime$}, and {\it i$^\prime$} bands, respectively, with an average of $\sim$20 TDI exposures collected each night. \par
For this work, we have used the commissioning phase data from the ILMT to develop and refine the astrometry pipeline and then used it to astrometrically calibrate the 4 cycles of data obtained till now.
%%%%%%%%%%%%%%%%%%%%%
\subsection{The Astrometric calibrators from the Gaia DR3}
\label{sec:gaia_data}
The European Space Agency’s Gaia mission \citep{2016A&A...595A...1G} has been scanning the entire sky at unprecedentedly high angular resolution. The Gaia DR3 \citep{2023A&A...674A...1G} provides celestial positions, proper motions, parallaxes, and broad band photometry for around 1.5 billion objects in the sky. For our analysis, we used the astrometric standard candidates identified by \citet{2022JAI....1140001D}, specifically constructed for ILMT using the Gaia DR3 data. The selected sample includes all Gaia DR3 sources within the declination range $28.9239^\circ$ to $29.8238^\circ$ (J2000), corresponding to the ILMT survey strip after accounting for Earth’s precession, and applying cuts for low proper motion ($\mu < 20$ mas/yr) and parallax ($\delta < 10$ mas), and further choosing sources with astrometric excess noise significance ($D < 2$), which quantifies the disagreement between the observation and the best-fitting astrometric model adopted by Gaia. The final catalog contains approximately $5.46 \times 10^6$ Gaia sources suitable for astrometric calibration of the ILMT fields. \par
%%%%%%%%%%%%%%%%%%%%%%%%%%%%
%%%%%%%%%%%%%%%%%%%%%%%%%%%%%
\section{Methodology}
\label{sec:methodology}

Our astrometric calibration proceeds in four steps: an initial plate solution of two sub-regions of each frame, crossmatching of the detected sources with the Gaia catalogue, propagation of the Gaia coordinates to the observation epoch, and a fit of transformation equations between the CCD pixel coordinates and the celestial coordinates. This is followed by updating the WCS in the fits header (see Section~\ref{sec:wcs_update}). We describe each step in turn.

Each ILMT TDI frame consists of $4096 \times 36864$ pixels, with 36864 pixels along the RA scan direction and 4096 pixels along the DEC direction, covering a sky area of $3.35^\circ$ in RA and $0.41^\circ$ in DEC. The first 4096 pixels of the raw frame in the RA direction are discarded, as they are affected by TDI ramping (see Figure~\ref{fig:tdi_ramping}). From each frame, we extract two chunks of $2048 \times 2048$ pixels each, one at the beginning and one at the end of the frame. 

We obtain an initial plate solution for the sources detected in the two chunks using the plate-solving engine \texttt{astrometry.net} \citep{2010AJ....139.1782L}, which yields approximate celestial coordinates with a precision of $\sim$0.5 arcsec. We note that plate solving the full TDI frame instead is computationally expensive, sometimes failing to converge within the allotted solve time, and yields larger astrometric offsets even when successful, as discussed in detail in Section~\ref{sec:comparison_section}.
These sources are then crossmatched, with a matching tolerance of 2 arcsec, against our catalogue of astrometric calibrators drawn from the Gaia survey (as discussed in Section~\ref{sec:gaia_data}), providing accurate Gaia coordinates for each matched source. To avoid contaminated astrometry, we reject all sources that have a companion within 2 arcsec.

The Gaia coordinates of the crossmatched calibrators are then converted to the observation epoch using the \texttt{astropy} package\footnote{\url{https://www.astropy.org}} \citep{2013A&A...558A..33A}, applying corrections for precession, nutation, and the aberration of light. 

In the presence of the TDI optical corrector, the stellar trajectories in the focal plane of the ILMT are nearly linear \citep{2024BSRSL..93..863N}. We therefore fit the following transformation equations between the CCD pixel coordinates and the epoch-converted celestial coordinates of the calibrators:

\begin{equation}
    \mathrm{RA} = f_{1} + f_{2}\,(x - x_{0}) + f_{3}\,(y - y_{0}),
    \label{eq:transformation_equations1_ILMT}
\end{equation}
\begin{equation}
    \mathrm{DEC} = g_{1} + g_{2}\,(y - y_{0}) + g_{3}\,(x - x_{0}) + g_{4}\,(x - x_{0})^{2},
    \label{eq:transformation_equations2_ILMT}
\end{equation}
where RA and DEC are the source coordinates at the observation epoch; $x$ and $y$ are the corresponding CCD pixel coordinates, with RA varying primarily along the $y$-direction and DEC along the $x$-direction; $x_{0}$ and $y_{0}$ are the central pixel positions; and $f_{1}, f_{2}, f_{3}, g_{1}, g_{2}, g_{3}, g_{4}$ are free parameters. The quadratic term in Equation~\ref{eq:transformation_equations2_ILMT} is required to account for the field distortion introduced by the telescope and corrector optics, as discussed in Section~\ref{sec:comparison_section}. 
The calibrators from the two chunks are combined and fitted simultaneously in a single global fit per frame, performed through non-linear least-squares minimisation using the \texttt{LMFIT} package \citep{2016ascl.soft06014N}, with the two equations solved together.

\begin{figure*}
   \centering
   \includegraphics[width=0.95\linewidth]{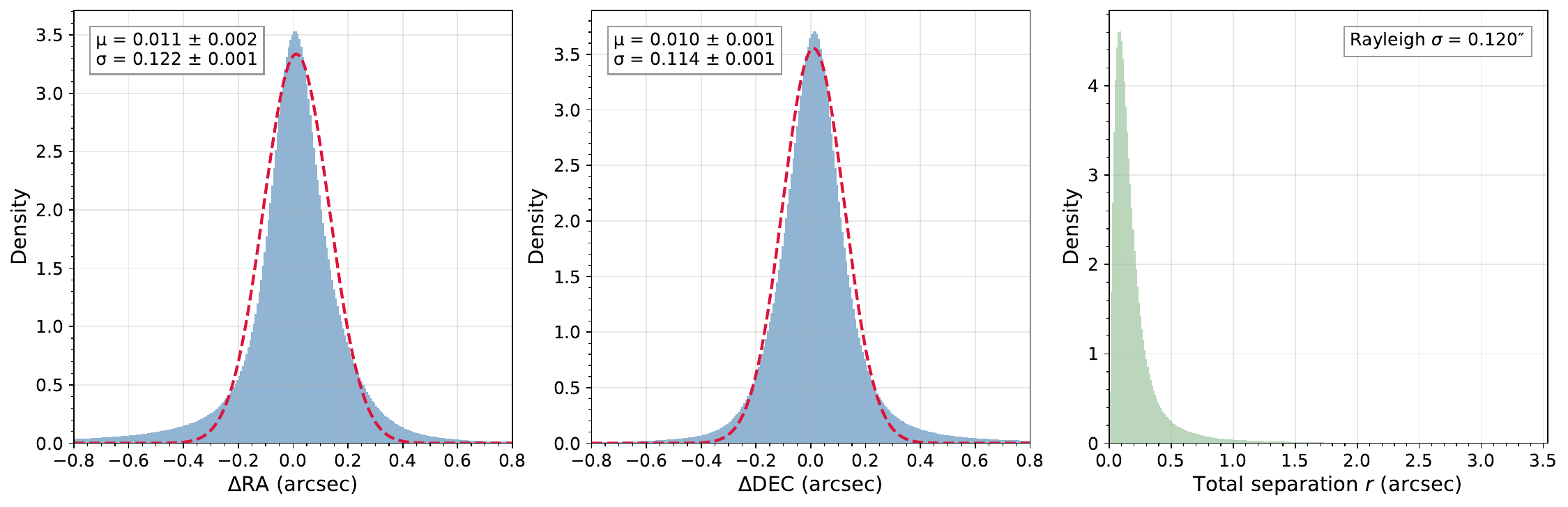}      \caption{Astrometric offsets from our custom astrometry pipeline for the data obtained with the ILMT during the cycles 1-4, compared with Gaia positions. The left plot shows the distribution of the offsets in the RA, whereas the center plot shows the offsets in DEC. The right most subplot shows the distribution in the total angular separation between our astrometric coordinates and the Gaia coordinates. The Rayleigh scale parameter $\sigma$ was estimated from the median of the total angular separation distribution using the analytic relation $\sigma = \mathrm{median}(r)/\sqrt{2\ln 2}$.}
         \label{fig:result_cycle3}
   \end{figure*}

These transformation relations yield sub-arcsec astrometric precision in the measured celestial positions of the detected sources, as shown in Figure~\ref{fig:result_cycle3}, which displays the distribution of the positional differences, relative to Gaia, of the sources detected in all TDI frames obtained with the ILMT during cycles 1--4. The median of the per-frame standard deviations of these differences is $\sim$0.12 arcsec in RA and $\sim$0.11 arcsec in DEC, while the mean offsets are $\sim$0.01 arcsec in both coordinates, indicating that the calibration is essentially free of systematic bias.

%%%%%%%%%%%%%%%%%%
%%%%%%%%%%%%%%%%%%%%%%%%%%%%%
\subsection{Updating the WCS information in the fits header}
\label{sec:wcs_update}
The transformation relations described in Equations \ref{eq:transformation_equations1_ILMT} and \ref{eq:transformation_equations2_ILMT} convert the CCD pixel coordinates to celestial coordinates (RA and DEC), but are not compatible with the FITS WCS convention \citep{fits_wcs_1, fits_wcs_2}, and hence not directly readable by established FITS parsing software like \texttt{wcslib\footnote{\url{https://www.atnf.csiro.au/computing/software/wcs/index.html}}, astropy} and \texttt{ds9\footnote{\url{https://sites.google.com/cfa.harvard.edu/saoimageds9}}}.\\

The FITS WCS convention establishes three steps to convert pixel coordinates $(x, y)$ to celestial world coordinates $(\alpha, \delta)$:
\begin{enumerate}
    \item Conversion of pixel coordinates $(x, y)$ to projection plane coordinates $(\mathcal{X}, \mathcal{Y})$ by means of a linear transformation (accounting for translation, rotation, skewness and scaling). This is governed by the \texttt{CRPIX}, and \texttt{CD} (or alternatively the \texttt{PC \& CDELT}) keywords in the FITS header.
    \item Conversion of projection plane coordinates $(\mathcal{X}, \mathcal{Y})$ to native spherical coordinates $(\phi, \theta)$ by reversing the assumed projection. This is governed by the \texttt{CTYPE} and \texttt{PV} keywords in the FITS header.
    \item Lastly, the conversion of native spherical coordinates $(\phi, \theta)$ to celestial coordinates $(\alpha, \delta)$ by means of rotation of this coordinate frame. This is governed by the \texttt{CRVAL} keyword in the FITS header.
\end{enumerate}

Since the ILMT images correspond to a strip of sky, the most natural projection to map the CCD coordinates to spherical coordinates is a cylindrical projection (in particular, the \texttt{Plate Carr\'ee} or \texttt{CAR} projection as described in \citealp{fits_wcs_2}).
The \texttt{CAR} projection relates $(\mathcal{X}, \mathcal{Y})$ to $(\phi, \theta)$ simply as $(\phi - \phi_0) = \mathcal{X}$ and $(\theta - \theta_0) = \mathcal{Y}$, where $\phi_0$ and $\theta_0$ are the native spherical coordinates of the reference pixel, defined by the \texttt{PV} keywords. In our case, these native spherical coordinates can be directly mapped to celestial world coordinates, without introducing any rotation, i.e. $\phi = \alpha$ and $\theta = \delta$.
Also, from Equations \ref{eq:transformation_equations1_ILMT} and \ref{eq:transformation_equations2_ILMT}, we see that $f_1$ and $g_1$ are the celestial coordinates of the reference pixel, therefore, 
\begin{align}
    \alpha_0 = \phi_0 = f1 \;\;\; \& \;\;\; \delta_0 = \theta_0 = g1 \label{eq:reference_pixel_coordinates}
\end{align}
% This can be achieved if the reference pixel has the same native pixel coordinate ($\phi_0, \theta_0$) and celestial coordinates ($\alpha_0, \delta_0$).

Posing, $u =  (x - x_0)$ and $v = (y - y_0)$, Equations \ref{eq:transformation_equations1_ILMT} and \ref{eq:transformation_equations2_ILMT} can be rewritten as 
\begin{align}
    (\alpha - \alpha_0)  &= (\phi - \phi_0) = \mathcal{X} = u \times f_2 + v \times f_3 + u^2 \times f_4 \label{eq:transform_eq_3}\\
    (\delta - \delta_0)  &= (\theta - \theta_0) = \mathcal{Y} =  u \times g_3 + v \times g_2 + u^2 \times g_4 \label{eq:transform_eq_4}
\end{align}
with $f_4 = 0$. This can be arranged as
\begin{align}
\begin{pmatrix}
    \mathcal{X} \\ \mathcal{Y}
\end{pmatrix}
=
\begin{pmatrix}
    f_2 & f_3 \\
    g_3 & g_2
\end{pmatrix}
\begin{pmatrix}
    u \\
    v 
\end{pmatrix}
+
\begin{pmatrix}
    f_4 \\
    g_4
\end{pmatrix} 
u^2 \label{eq:transform_matrix}
\end{align}

With Equations \ref{eq:transform_eq_3}, \ref{eq:transform_eq_4} and \ref{eq:transform_matrix}, we have solved step (ii) and (iii) of the FITS WCS convention. However, we notice that $(\mathcal{X}, \mathcal{Y})$ have a non-linear dependence on $(u, v)$, and this cannot be achieved by a linear matrix transformation. To faithfully represent these transformations we adopt the SIP convention formulated by \cite{fits_sip_convention}, which allows us to use polynomial distortion terms in the following form: 
\begin{align}
\begin{pmatrix}
    \mathcal{X} \\ \mathcal{Y}
\end{pmatrix}
=
\begin{pmatrix}
    CD_{11} & CD_{12} \\
    CD_{21} & CD_{22}
\end{pmatrix}
\begin{pmatrix}
    u + f(u, v) \\
    v + g(u, v)
\end{pmatrix} \nonumber 
\\
=
CD
\begin{pmatrix}
    u \\
    v 
\end{pmatrix}
+
CD
\begin{pmatrix}
    f(u, v) \\
    g(u, v)
\end{pmatrix} \label{eq:sip_convention}
\end{align}
where
\[
f(u, v) = A_{20} u^2 + A_{02} v^2 + A_{11} uv \mathrm{ \,\,, and}
\]
\[
g(u, v) = B_{20} u^2 + B_{02} v^2 + B_{11} uv \,,
\]
for a second order polynomial distortion, where A and B are distortion coefficients.

Comparing Equations \ref{eq:transform_matrix} and \ref{eq:sip_convention}, we see that $f_2, f_3, g_3, g_2$ are the elements of the $CD$ matrix, $CD_{11}, CD_{12}, CD_{21}, CD_{22}$, respectively. It can be shown that
\begin{align}
\begin{pmatrix}
    A_{20} \\
    B_{20} 
\end{pmatrix}
=
(CD)^{-1}
\begin{pmatrix}
    f_4 \\
    g_4 
\end{pmatrix} 
\label{eq:fg_to_ab}
\end{align}
With this, we can construct a FITS header, compatible with the WCS conventions, that encodes the relations described in Equations \ref{eq:transformation_equations1_ILMT} and \ref{eq:transformation_equations2_ILMT}. An excerpt from the header of a representative ILMT frame \footnote{The WCS information in the headers of ILMT frames will not be parsed correctly by \texttt{astropy} $< 5.2.3$ due to a bug in the software.}, which can be parsed with FITS WCS-compliant software, is shown below.\\

\parindent 0em
{\small
\begin{verbatim}
CTYPE1  = `RA---CAR-SIP' / CAR with SIP
CTYPE2  = `DEC--CAR-SIP' / CAR with SIP
CUNIT1  = `deg'
CRVAL1  = 106.1443849525412 / RA at ref pixel
CRPIX1  = 2048.0 / ref pixel x value
CUNIT2  = `deg'
CRVAL2  = 29.353276026966 / DEC at ref pixel
CRPIX2  = 18432.0 / ref pixel y value
CD1_1   = 8.1059953709469E-09
CD1_2   = 0.000104121431956795
CD2_1   = -9.0831859783511E-05
CD2_2   = 5.73083225800705E-10
EQUINOX = 2023.81853711935 / Observation epoch
RADESYS = `FK5'
PV1_0   = 1
PV1_1   = 106.1443849525412
PV1_2   = 29.353276026966
A_ORDER = 2 / poly order axis 1
A_2_0   = 4.55491917624246E-07 / distortion coeff
A_0_2   = 0.0 / distortion coeff
B_ORDER = 2 / poly order axis 2
B_2_0   = -3.1086044412880E-11 / distortion coeff
B_0_2   = 0.0 / distortion coeff
\end{verbatim}}

In this header, the \texttt{CTYPE} keyword describes the projection and also indicates that \texttt{SIP} distortions are used. The \texttt{CRPIX, CRVAL}, and \texttt{CD} keywords encode the values of $(x_0, y_0)$, $(\alpha_0, \delta_0)$, and $(f_2, f_3, g_3, g_2)$, respectively. The second and higher order \texttt{SIP} coefficients that have not been explicitly specified assume a default value of zero.
%%%%%%%%%%%%%%%%%%%%%%%%%%%%%%%%%%%%%%%%%%%%%%%%%%%%%%%%%%%%%%%
%%%%%%%%%%%%%%%%%%%%%%%%%%%%%%%%%%%%%%%%%%%%%%%%%%%%%%%%%%%%%%%%%%
\begin{figure}
\centering
\includegraphics[width=0.99\linewidth]{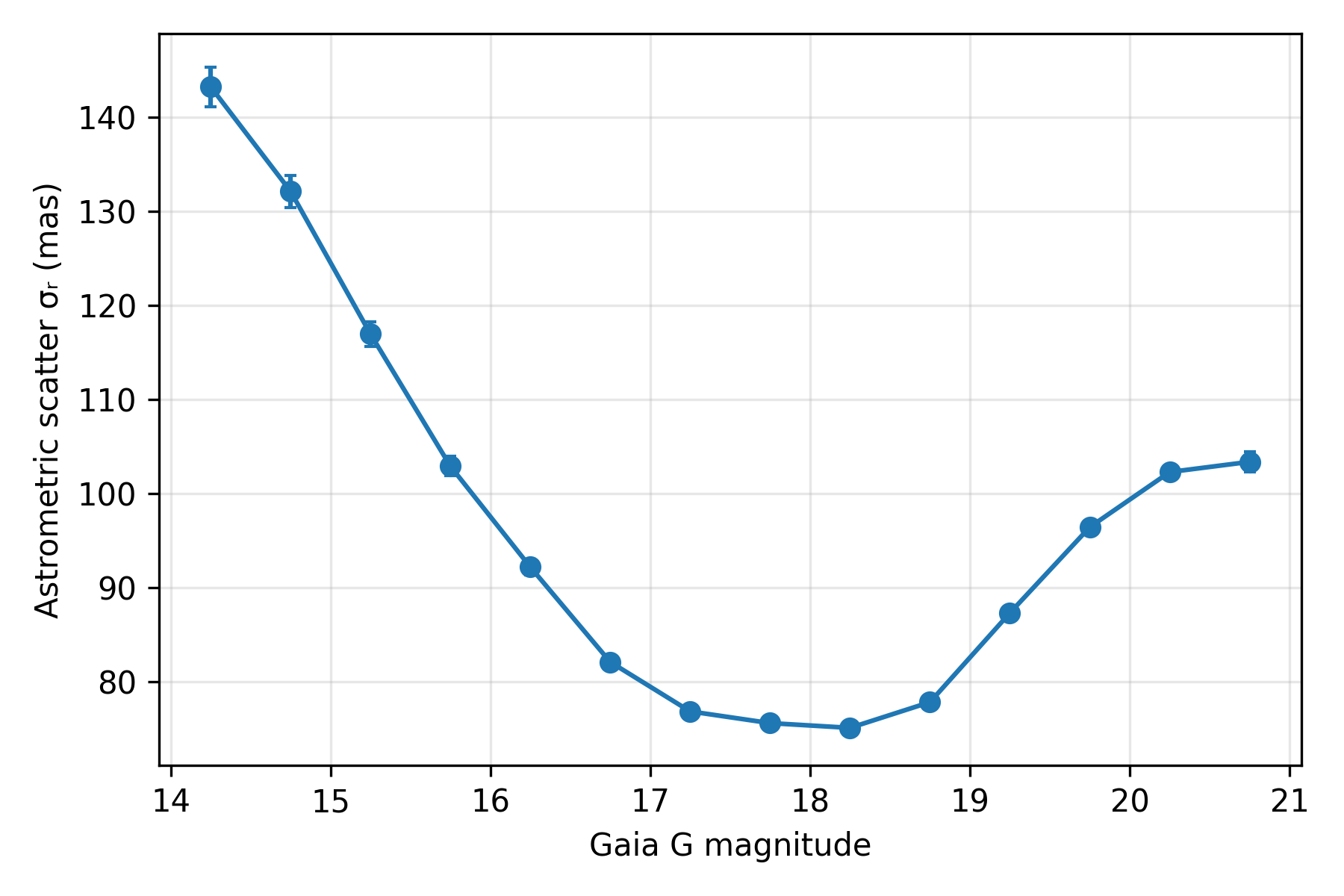}
  \caption{Astrometric scatter ($\sigma_{r}$) as a function of Gaia G magnitude for our fine astrometric solution. The scatter in each magnitude bin is estimated using a robust MAD-based statistic and is computed for all the stars detected with ILMT on 2024-02-07, in 26 individual frames (\textit{i} band). The precision improves from bright magnitudes toward an optimal range around G $\sim$16.5-18.5, and degrades again toward fainter magnitudes due to decreasing signal-to-noise.}
     \label{fig:result5}
\end{figure}  
%%%%%%%%%%%%%%%%%%%%%%%%%%%%%%%%%%%%%%%%%%%%%%%%%%%%%%%%%%%%%%%%%%
     \begin{figure*}
   \centering
   \includegraphics[width=0.95\linewidth]{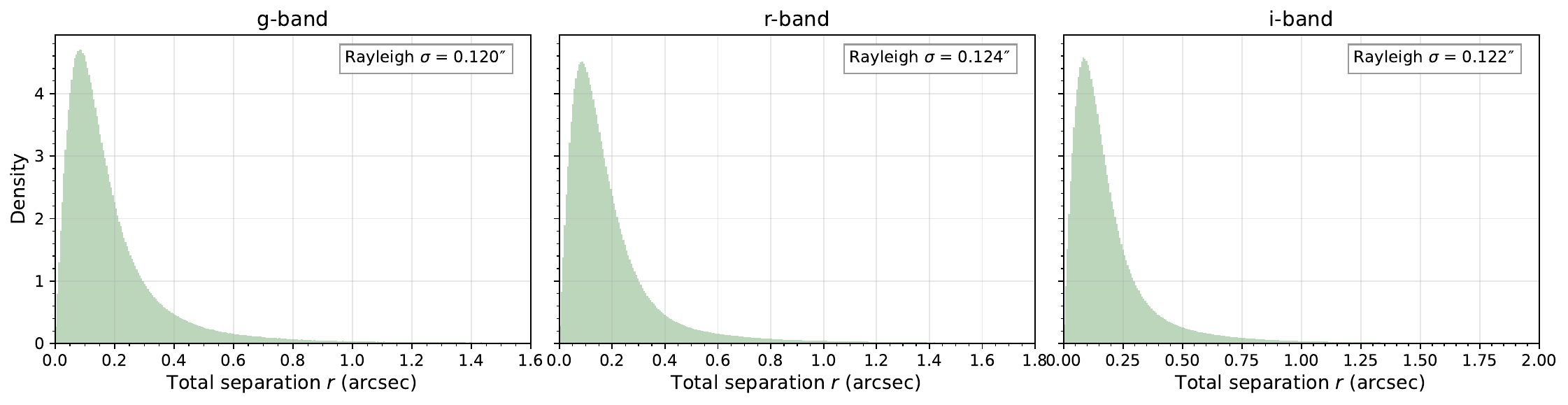}
      \caption{ The three subplots show the distribution in the total angular separation between our astrometric coordinates and the Gaia coordinates, for the three ILMT bands $g^{\prime}$, $r^{\prime}$ and $i^{\prime}$, respectively.}
         \label{fig:result_cycle3_bandwise}
   \end{figure*}
%%%%%%%%%%%%%%%%%%%%
%%%%%%%%%%%%%%%%%%%%
     \begin{figure}
   \centering
   \includegraphics[width=0.99\linewidth]{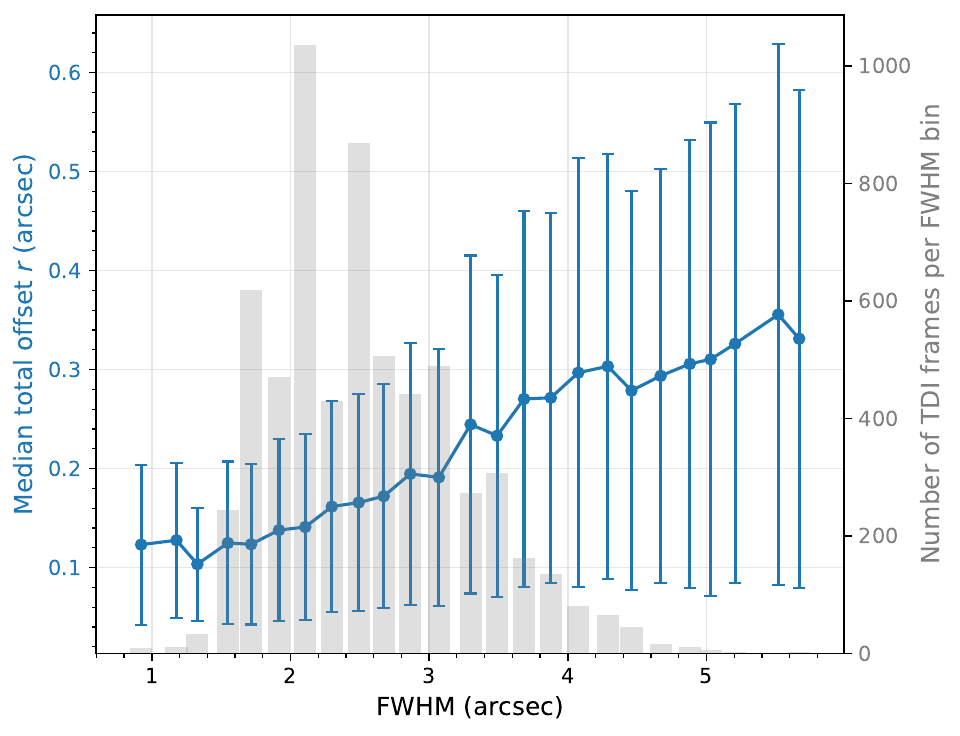}
      \caption{Median astrometric offset as a function of atmospheric seeing, for the full cycle-3 and cycle-4 data. The median total astrometric residual is shown as a function of the image FWHM. Blue circles show the median total residual per FWHM bin, computed from the median offsets of individual TDI frames. Vertical error bars represent the median absolute deviation (MAD) within each bin, providing a robust estimate of the frame-to-frame scatter. The gray histogram indicates the number of TDI frames contributing to each seeing bin.}
         \label{fig:seeing_variation}
   \end{figure}
%%%%%%%%%%%%%%%%%%%%

 \section{Results from the four ILMT cycles and comparison with alternate approach}
 \label{sec:results}
 %***************************
We have tested the astrometry pipeline on all the TDI frames obtained with the ILMT in the 4 cycles. This includes $\sim$347 nights of data with $\sim$20 TDI frames per night. The median offsets in the calculated RA-DEC as compared to the Gaia coordinates, for each frame are shown in Figure~\ref{fig:result_cycle3}. This indicates an astrometric accuracy of $\sim$0.12 arcsec in RA and $\sim$0.11 arcsec in DEC, considering all the detections in each frame, with a Rayleigh scale parameter in the total offset of $\sim$0.12 arcsec. Here, we further investigate the dependence of the astrometric performance on source magnitude, image quality (seeing), and compare results with another approach.\par
%%%%%%%%%%%%%%%%%%%%%%%%%%%%%
%%%%%%%%%%%%%%%%%%%%%%%%%%%%%%%%%%%%%%%%%%%%%%%%%%%%%%%%%%%%%%%%%%
\subsection{Astrometric precision as a function of magnitude}
The ILMT detects sources over a magnitude range of $\sim$14.5 - 21.5 mag (Ailawadhi et al., in preparation). Figure~\ref{fig:result5} shows the dependence of the astrometric precision on source brightness using our custom astrometric solution, for one night of observations obtained on 2024-02-07, covering 26 individual TDI frames. The measured scatter decreases from bright magnitudes toward an optimal regime around G $\sim$ 16.5–18.5, where the minimum scatter of $\sim$75$-$80 milli-arcsec is achieved. 
This corresponds to the regime in which centroiding accuracy is maximized for unsaturated, high signal-to-noise ratio (SNR) sources.
At fainter magnitudes, the precision gradually degrades as photon noise and background contributions become dominant. The smooth and monotonic behaviour across the full magnitude range, together with the absence of sharp discontinuities, suggests that the residual astrometric errors are largely governed by statistical uncertainties rather than by uncorrected systematic effects.
%%%%%%%%%%%%%%%%%%%%%%%%%%%%%%%%%%%%%%%%%%%%%%%%%%%%%%%%%%%%%%%%%%
\subsection{Bandwise accuracy and corelation with seeing}
%%%%%%%%%%%%%%%%%%%%%%%%%%%%%%%
We also evaluated the bandwise astrometric precision of our astrometry pipeline, using the Cycle-3 ILMT data. The distributions of total angular separations in the $g^{\prime}$, $r^{\prime}$, and $i^{\prime}$ bands exhibit very similar behaviour, with Rayleigh scale parameters of $\sim$0.12 arcsec in all three filters (see Figure~\ref{fig:result_cycle3_bandwise}). This close agreement indicates that the astrometric performance is largely independent of wavelength across the ILMT optical range, and that no significant band-dependent systematic offsets are present. The marginally smaller scale parameters in the $r$ and $i$ bands may reflect slightly improved centroiding stability at longer wavelengths, where atmospheric turbulence and chromatic effects are reduced. Overall, the near-identical distributions across filters demonstrate the internal consistency and robustness of the astrometric calibration procedure.

%%%%%%%%%%%%%%%%%%%%
Further, the seeing across the ILMT frames was found to vary between $\sim$1.5 arcsec and $\sim$3.5 arcsec in most cases, with a few frames reaching values as high as $\sim$5 arcsec. Hence, we also investigated the astrometric offsets as a function of the varying seeing for the full cycle-3 and cycle-4 ILMT data. 
As shown in Fig.~\ref{fig:seeing_variation}, the astrometric performance exhibits a moderate but systematic degradation with increasing atmospheric seeing. The median total astrometric offset increases from $\lesssim 0.10$ arcsec under good conditions (FWHM $\lesssim 1.5$ arcsec) to $\lesssim 0.30$ arcsec for poorer seeing (FWHM $\sim 4$ arcsec). This behaviour is consistent with expectations, as centroiding precision scales with the PSF width and therefore degrades as the image quality worsens.
The increased scatter observed at larger FWHM values reflects both the intrinsic broadening of stellar profiles and the reduced number of frames contributing to those bins. Importantly, the majority of the observations fall within the 2–3 arcsec seeing regime, where the astrometric solution remains stable and well constrained, with median offsets below $\sim 0.15$ arcsec.
%\end{itemize}
%%%%%%%%%%%%%%%%%%%%%%%%%%%%%%%%%%%%%%%%%%%%%%%%%%%%%%%%%%%%%%%%%%
\subsection{Comparing the results with alternate approaches}
\label{sec:comparison_section}
%%%%%%%%%%%%%%%%%%%%%%%%%%%%%%%%%%%%%%%%%%%%%%%%%%%%%%%%%%%%%%%%%%
    \begin{figure*}
   \centering
    \includegraphics[width=0.9\linewidth]{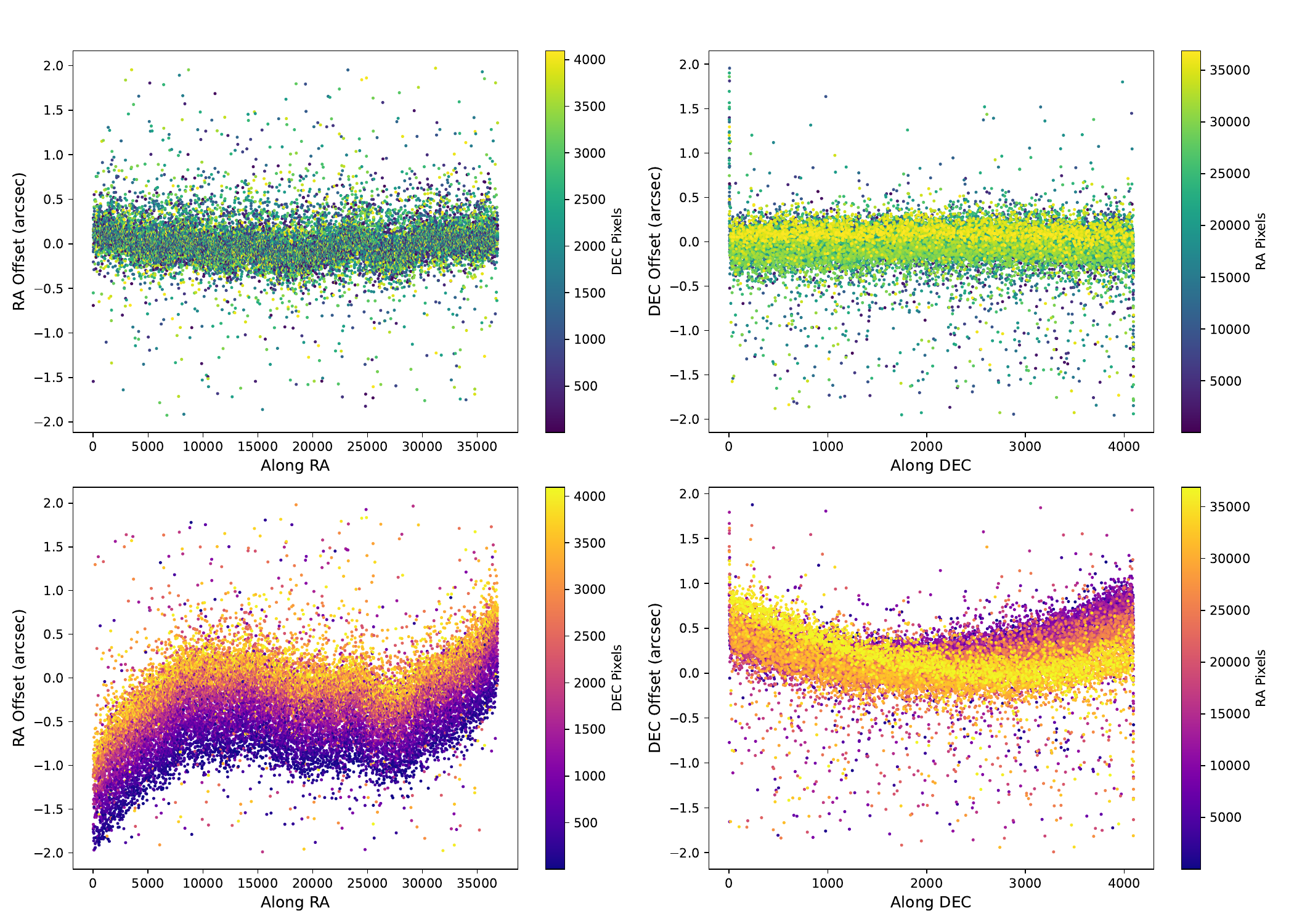}
      \caption{Astrometric offsets in one of the TDI frames observed with ILMT on 2024-02-07. The top panel shows the offsets in RA (/DEC) as a function of pixels along RA (/DEC), from our fine astrometry. The pixels along DEC (/RA) are color coded to show variation along the DEC (/RA) direction. The bottom panels show the same using `astrometry.net' for the full frame astrometry.}
         \label{fig:result3}
   \end{figure*}   
%%%%%%%%%%%%%%%%%%%%%%%%%%%%%%%%%%%%%%%%%%%%%%%%%%%%%%%%%%%%%%%%%%
%%%%%%%%%%%%%%%%%%%%%%%%%%%%%%%%%%%%%%%%%%%%%%%%%%%%%%%%%%%%%%%%%%
\begin{figure*}
\centering
\includegraphics[width=0.99\linewidth,height=0.35\textheight,keepaspectratio]{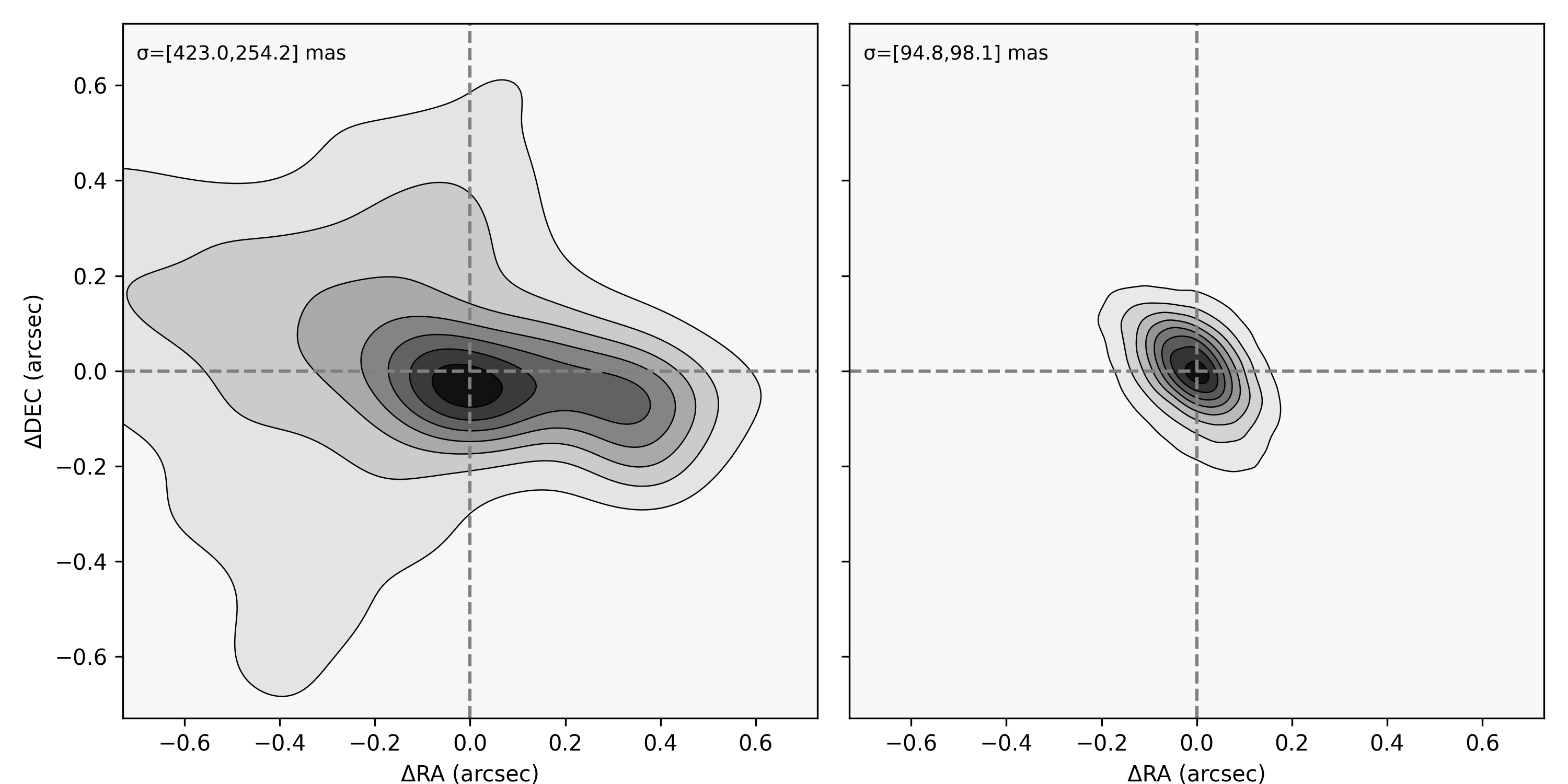}
  \caption{Two-dimensional residual distributions of astrometric offsets with respect to Gaia DR3 for all the objects ($17 < i_{mag} < 19$) detected with ILMT on 2024-02-07, in 26 individual full TDI frames, observed in $i-$band. The left panel shows the initial astrometric solution from \texttt{astrometry.net}, while the right panel shows the refined astrometric solution using our custom pipeline. Contours indicate kernel-density estimates of the residual distribution in $\Delta$RA and $\Delta$DEC (arcsec), with dashed lines marking zero offset. The refined solution exhibits a compact, nearly isotropic core centered close to zero, with a robust per-coordinate scatter of $\sim$100 mas.}
     \label{fig:result4}
\end{figure*}  
%%%%%%%%%%%%%%%%%%%%%%%%%%%%%%%%%%%%%%%%%%%%%%%%%%%%%%%%%%%%%%%%%
We also tested astrometric calibration of full ILMT TDI frames (4096 $\times$ 36864) using the \texttt{astrometry.net} software. 
The resulting astrometric solutions from \texttt{astrometry.net} were found to be reasonably accurate, with positional residuals typically within 1 arcsec (see Figure~\ref{fig:result3} bottom panel, and Figure~\ref{fig:result4}). 
However, as apparent in Figure~\ref{fig:result3}, the residuals are larger and spatially coherent, with systematic curvature patterns reaching amplitudes of up to $\sim$1–1.5 arcsec. These residual structures indicate that generic low-order distortion models are insufficient to fully describe the wide-field TDI geometry of ILMT. In contrast, our physically motivated instrument-specific transformation model (Eq.~\ref{eq:transformation_equations1_ILMT} and Eq.~\ref{eq:transformation_equations2_ILMT}), which incorporates a second-order correction in the DEC direction, effectively suppresses these large-scale trends, resulting in spatially uniform residuals and reduced scatter, as illustrated in Figure~\ref{fig:result4}. \par
Overall, our pipeline achieves a factor of $\sim$3$-$4 improvement in the astrometric accuracy, and this superior performance of the custom model is due to its closer alignment with the residual distortion structure of a TDI system equipped with an optical corrector. The TDI optical corrector is designed to compensate for the dominant projection curvature and field-dependent distortion along the scan direction, substantially reducing higher-order geometric effects across the full drift length \citep{2024BSRSL..93..863N}. As a result, the remaining distortions are smooth, low-order, and largely one-dimensional. The adopted transformations explicitly reflect this behavior by separating along-scan and cross-scan terms and including only a quadratic term in the DEC direction. This additional second-order term empirically accounts for a mild residual asymmetry in the optical system, potentially associated with small alignment offsets within the TDI corrector. The resulting low-order, physically motivated model provides a more faithful mapping between detector and sky coordinates for corrected TDI data. \par
Furthermore, for a small but significant fraction of TDI frames, the \texttt{astrometry.net} processing fails due to time-out issues. This is most likely because of the very wide size of ILMT images ($\sim$3.3$^\circ$ in RA), combined with image distortion, vignetting, or warping, which increase the complexity of the star-matching algorithm. Since \texttt{astrometry.net} relies on matching geometric star patterns (triangles or quads) against precomputed index catalogs, for very wide and elongated fields with high stellar densities, the number of plausible candidate matches can increase substantially, leading to higher computational cost during hypothesis generation and verification. In addition, the long and narrow geometry of ILMT TDI images may lead to nearly collinear or poorly conditioned stellar quads, making the geometric hashes less distinctive and further complicating the matching process. In practice, this can result in long solve times or time-out failures for some frames. Hence, we found that our approach of extracting two small chunks from the full TDI image and calibrating the entire frame using linear transformation relations with second-order corrections is both optimal and efficient for handling large ILMT frames.
%%%%%%%%%%%%%%%%%%%%%%%%%%%%
%______________________________________________________________
    
\section{Summary and Conclusions}
\label{sec:discussion}
The ILMT is a survey telescope dedicated to time-domain astronomy, covering $\sim$120 square degrees of the sky in {\it g$^\prime$}, {\it r$^\prime$}, and {\it i$^\prime$} optical bands. We have developed a robust astrometric calibration pipeline for the ILMT data obtained in TDI mode. The pipeline employs a near-linear transformation model from pixel to world coordinates, augmented with a second order correction for the asymmetrical distortion introduced by the TDI mode of observations, consisting of hyperbolic stellar trajectories in the focal plane corrected at best by a well designed optical corrector \citep{1998PASP..110.1081H, 2024BSRSL..93..863N}. 

Applying the pipeline to 347 nights of data from the first four observing cycles (2022--2025) of ILMT, we demonstrate a typical astrometric accuracy of $\sim$100 milliarcseconds (mas), improving to $\sim70$--$80$ mas for moderately bright stars ($G \sim 16.5$--$18.5$). 
Comparison to the widely used plate-solving engine \texttt{astrometry.net}, our instrument specific approach yields a factor of $\sim$3--4 lower positional residuals against Gaia DR3 and effectively eliminates the systematic trends present in the generic solutions. This improved performance arises from the instrument-specific and physically motivated modeling adopted in our pipeline. By explicitly accounting for the characteristic TDI scan geometry and incorporating a minimal second-order correction in the cross-scan direction, the transformation more accurately captures the residual low-order distortion structure of ILMT data. 

The pipeline is integrated into the routine ILMT data-processing workflow, and the astrometrically calibrated data products from cycles 1--4 are publicly available. The demonstrated accuracy and multi-year stability establish a validated framework for precision astrometry with zenith-pointing TDI surveys, and provide the astrometric foundation for forthcoming ILMT science, including transient localization, variability characterization, and long-term positional monitoring of Solar System and Galactic objects.

%%%%%%%%%%%%%%%%%%%%%%%%%%%%%
\begin{acknowledgements}
We thank the anonymous referee for carefully reviewing our manuscript and providing insightful comments. The 4-m International Liquid Mirror Telescope (ILMT) project results from a collaboration between the Institute of Astrophysics and Geophysics (University of Liège, Belgium), the Universities of British Columbia, Laval, Montreal, Toronto, Victoria and York University, and Aryabhatta Research Institute of Observational SciencES (ARIES, India). The authors thank Hitesh Kumar, Himanshu Rawat, Khushal Singh, Ankit Bisht, Nikhil Dharkiya, and other observing staff for their assistance at the 4-m ILMT. The team acknowledges the contributions of ARIES’s past and present scientific, engineering and administrative members in the realisation of the ILMT project. 
This work has made use of data from the European Space Agency (ESA) mission {\it Gaia} (\url{https://www.cosmos.esa.int/gaia}), processed by the {\it Gaia} Data Processing and Analysis Consortium (DPAC, \url{https://www.cosmos.esa.int/web/gaia/dpac/consortium}). Funding for the DPAC has been provided by national institutions, in particular the institutions participating in the {\it Gaia} Multilateral Agreement.
VN is supported by Beijing Natural Science Foundation (Grant No. IS25004). JS wishes to thank Service Public Wallonie, F.R.S.-FNRS (Belgium) and the University of Liège, Belgium for funding the construction of the ILMT. PH acknowledges financial support from the Natural Sciences and Engineering Research Council of Canada, RGPIN-2019-04369. PH and JS thank ARIES for hospitality during their visits to Devasthal. B.A. acknowledges the Council of Scientific \& Industrial Research (CSIR) fellowship award (09/948(0005)/2020-EMR-I) for this work. M.D. acknowledges Innovation in Science Pursuit for Inspired Research (INSPIRE) fellowship award (DST/INSPIRE Fellowship/2020/IF200251) for this work. JS and KM acknowledge the assistance received from the Anusandhan National Research Foundation (ANRF, SERB- 762 VAJRA Faculty Scheme, India). B.K. acknowledges the``Special Project for High-End Foreign Experts,” Xingdian Funding from Yunnan Province. 
\end{acknowledgements}

\begin{contribution}
%%This section gives authors the space to recognize author contributions. The text inside this environment is NOT counted towards the total word quanta. At a minimum, manuscripts are expected to include this text:

This work results from a long term collaboration where all authors have contributed equally.

%% But authors are expected to provide more specific details, e.g. 
%%
%%SC was responsible for writing and submitting the manuscript.
%%WWM came up with the initial research concept and edited the manuscript.
%%OTS obtained the funding and edited the manuscript.
%%EBF provided the formal analysis and validation. He also edited the manuscript.
%%GEH Supervised the undergraduates, wrote the software and administers the project github and Zenodo repositories.
%%
%% Authors can use the Contributor Role Taxonomy (CRediT) at
%% for ideas on how write a good statement tailored to their needs.

\end{contribution}

% \facilities{HST(STIS), Swift(XRT and UVOT), AAVSO, CTIO:1.3m, CTIO:1.5m, CXO}

% \software{astropy \citep{2013A&A...558A..33A,2018AJ....156..123A,2022ApJ...935..167A},  
          % Cloudy \citep{2013RMxAA..49..137F}, 
          % Source Extractor \citep{1996A&AS..117..393B}
          % }


\bibliography{PASPsample701}{}

@ARTICLE{2025A&A...694A..80S,
       author = {{Surdej}, J. and {Hickson}, P. and {Misra}, K. and {Banerjee}, D. and {Ailawadhi}, B. and {Akhunov}, T. and {Borra}, E. and {Dubey}, M. and {Dukiya}, N. and {Filali}, S. and {Hellemeier}, J. and {Kharayat}, M. and {Kumar}, B. and {Kumar}, H. and {Kumar}, M. and {Kumar}, T.~S. and {Kumari}, P. and {Negi}, V. and {Pospieszalska-Surdej}, A. and {Prabhavu}, S. and {Pradhan}, B. and {Pranshu}, K. and {Rawat}, H. and {Reddy}, B.~K. and {Sasidharan Pillai}, A. and {Singh}, K. and {Tremblay}, S. and {Turakhia}, S. and {Vijay}, S.},
        title = "{The 4 m International Liquid Mirror Telescope: Construction, operation, and science}",
      journal = {\aap},
         year = 2025,
        month = feb,
       volume = {694},
          eid = {A80},
        pages = {A80},
          doi = {10.1051/0004-6361/202452667},
archivePrefix = {arXiv},
       eprint = {2502.00564},
 primaryClass = {astro-ph.IM},
       adsurl = {https://ui.adsabs.harvard.edu/abs/2025A&A...694A..80S}
}

@ARTICLE{2022JAI....1140001D,
       author = {{Dukiya}, Naveen and {Misra}, Kuntal and {Pradhan}, Bikram and {Negi}, Vibhore and {Ailawadhi}, Bhavya and {Kumar}, Brajesh and {Hickson}, Paul and {Surdej}, Jean},
        title = "{Astrometric and Photometric Standard Candidates for the Upcoming 4-m International Liquid Mirror Telescope Survey}",
      journal = {Journal of Astronomical Instrumentation},
         year = 2022,
        month = jan,
       volume = {11},
       number = {4},
          eid = {2240001},
        pages = {2240001},
          doi = {10.1142/S2251171722400013},
archivePrefix = {arXiv},
       eprint = {2210.06473},
 primaryClass = {astro-ph.IM},
       adsurl = {https://ui.adsabs.harvard.edu/abs/2022JAI....1140001D}
}

@ARTICLE{2024BSRSL..93.1054S,
       author = {{Surdej}, Jean and {Ailawadhi}, Bhavya and {Akhunov}, Talat and {Borra}, Ermanno and {Dubey}, Monalisa and {Dukiya}, Naveen and {Fu}, Jiuyang and {Grewal}, Baldeep and {Hickson}, Paul and {Kumar}, Brajesh and {Misra}, Kuntal and {Negi}, Vibhore and {Pospieszalska-Surdej}, Anna and {Pranshu}, Kumar and {Sun}, Ethen},
        title = "{The 4m International Liquid Mirror Telescope: a Brief History and Some Preliminary Scientific Results}",
      journal = {Bulletin de la Societe Royale des Sciences de Liege},
         year = 2024,
        month = jun,
       volume = {93},
       number = {2},
        pages = {1054-1068},
          doi = {10.25518/0037-9565.11961},
archivePrefix = {arXiv},
       eprint = {2311.05623},
 primaryClass = {astro-ph.IM},
       adsurl = {https://ui.adsabs.harvard.edu/abs/2024BSRSL..93.1054S}
}

@ARTICLE{2024BSRSL..93..863N,
       author = {{Negi}, Vibhore and {Ailawadhi}, Bhavya and {Akhunov}, Talat and {Borra}, Ermanno and {Dubey}, Monalisa and {Dukiya}, Naveen and {Fu}, Jiuyang and {Grewal}, Baldeep and {Hickson}, Paul and {Kumar}, Brajesh and {Misra}, Kuntal and {Pranshu}, Kumar and {Sun}, Ethen and {Surdej}, Jean},
        title = "{Necessity of a TDI Optical Corrector for ILMT Observations}",
      journal = {Bulletin de la Societe Royale des Sciences de Liege},
         year = 2024,
        month = jun,
       volume = {93},
       number = {2},
        pages = {863-871},
          doi = {10.25518/0037-9565.11904},
archivePrefix = {arXiv},
       eprint = {2311.04712},
 primaryClass = {astro-ph.IM},
       adsurl = {https://ui.adsabs.harvard.edu/abs/2024BSRSL..93..863N}
}

@ARTICLE{2024BSRSL..93..872M,
       author = {{Misra}, Kuntal and {Ailawadhi}, Bhavya and {Akhunov}, Kuntal and {Borra}, Ermanno and {Dubey}, Monalisa and {Dukiya}, Naveen and {Fu}, Jiuyang and {Grewal}, Baldeep and {Hickson}, Paul and {Kumar}, Brajesh and {Negi}, Vibhore and {Pranshu}, Kumar and {Sun}, Ethen and {Surdej}, Jean},
        title = "{Accessibility of the ILMT Survey Data}",
      journal = {Bulletin de la Societe Royale des Sciences de Liege},
         year = 2024,
        month = jun,
       volume = {93},
       number = {2},
        pages = {872-879},
          doi = {10.25518/0037-9565.11908},
archivePrefix = {arXiv},
       eprint = {2311.04717},
 primaryClass = {astro-ph.IM},
       adsurl = {https://ui.adsabs.harvard.edu/abs/2024BSRSL..93..872M}
}

@ARTICLE{2024BSRSL..93..820A,
       author = {{Ailawadhi}, Bhavya and {Akhunov}, Talat and {Borra}, Ermanno and {Dubey}, Monalisa and {Dukiya}, Naveen and {Fu}, Jiuyang and {Grewal}, Baldeep and {Hickson}, Paul and {Kumar}, Brajesh and {Misra}, Kuntal and {Negi}, Vibhore and {Pranshu}, Kumar and {Sun}, Ethen and {Surdej}, Jean},
        title = "{An Automated Photometric Pipeline for the ILMT data}",
      journal = {Bulletin de la Societe Royale des Sciences de Liege},
         year = 2024,
        month = jun,
       volume = {93},
       number = {2},
        pages = {820-827},
          doi = {10.25518/0037-9565.11892},
archivePrefix = {arXiv},
       eprint = {2311.04713},
 primaryClass = {astro-ph.IM},
       adsurl = {https://ui.adsabs.harvard.edu/abs/2024BSRSL..93..820A}
}

@ARTICLE{2025MNRAS.538..133P,
       author = {{Pranshu}, Kumar and {Misra}, Kuntal and {Ailawadhi}, Bhavya and {Dubey}, Monalisa and {Dukiya}, Naveen and {Filali}, Sara and {Hickson}, Paul and {Kumar}, Brajesh and {Negi}, Vibhore and {Surdej}, Jean},
        title = "{PYLMT : a transient detection pipeline for the 4-m International Liquid Mirror Telescope}",
      journal = {\mnras},
         year = 2025,
        month = mar,
       volume = {538},
       number = {1},
        pages = {133-152},
          doi = {10.1093/mnras/staf206},
archivePrefix = {arXiv},
       eprint = {2502.00556},
 primaryClass = {astro-ph.IM},
       adsurl = {https://ui.adsabs.harvard.edu/abs/2025MNRAS.538..133P}
}

@ARTICLE{2018BSRSL..87...68S,
       author = {{Surdej}, Jean and {Hickson}, Paul and {Borra}, Hermanno and {Swings}, Jean-Pierre and {Habraken}, Serge and {Akhunov}, Talat and {Bartczak}, Przemyslaw and {Chand}, Hum and {De Becker}, Micha{\"e}l and {Delchambre}, Ludovic and {Finet}, Fran{\c{c}}ois and {Kumar}, Brajesh and {Pandey}, Anil and {Pospieszalska}, Anna and {Pradhan}, Bikram and {Sagar}, Ram and {Wertz}, Olivier and {De Cat}, Peter and {Denis}, Stefan and {de Ville}, Jonathan and {Jaiswar}, Mukesh Kumar and {Lampens}, Patricia and {Nanjappa}, Nandish and {Tortolani}, Jean-Marc},
        title = "{The 4-m International Liquid Mirror Telescope}",
      journal = {Bulletin de la Societe Royale des Sciences de Liege},
         year = 2018,
        month = apr,
       volume = {87},
        pages = {68-79},
       adsurl = {https://ui.adsabs.harvard.edu/abs/2018BSRSL..87...68S}
}

@ARTICLE{2016arXiv161205560C,
       author = {{Chambers}, K.~C. and {Magnier}, E.~A. and {Metcalfe}, N. and {Flewelling}, H.~A. and {Huber}, M.~E. and {Waters}, C.~Z. and {Denneau}, L. and {Draper}, P.~W. and {Farrow}, D. and {Finkbeiner}, D.~P. and {Holmberg}, C. and {Koppenhoefer}, J. and {Price}, P.~A. and {Rest}, A. and {Saglia}, R.~P. and {Schlafly}, E.~F. and {Smartt}, S.~J. and {Sweeney}, W. and {Wainscoat}, R.~J. and {Burgett}, W.~S. and {Chastel}, S. and {Grav}, T. and {Heasley}, J.~N. and {Hodapp}, K.~W. and {Jedicke}, R. and {Kaiser}, N. and {Kudritzki}, R. -P. and {Luppino}, G.~A. and {Lupton}, R.~H. and {Monet}, D.~G. and {Morgan}, J.~S. and {Onaka}, P.~M. and {Shiao}, B. and {Stubbs}, C.~W. and {Tonry}, J.~L. and {White}, R. and {Ba{\~n}ados}, E. and {Bell}, E.~F. and {Bender}, R. and {Bernard}, E.~J. and {Boegner}, M. and {Boffi}, F. and {Botticella}, M.~T. and {Calamida}, A. and {Casertano}, S. and {Chen}, W. -P. and {Chen}, X. and {Cole}, S. and {Deacon}, N. and {Frenk}, C. and {Fitzsimmons}, A. and {Gezari}, S. and {Gibbs}, V. and {Goessl}, C. and {Goggia}, T. and {Gourgue}, R. and {Goldman}, B. and {Grant}, P. and {Grebel}, E.~K. and {Hambly}, N.~C. and {Hasinger}, G. and {Heavens}, A.~F. and {Heckman}, T.~M. and {Henderson}, R. and {Henning}, T. and {Holman}, M. and {Hopp}, U. and {Ip}, W. -H. and {Isani}, S. and {Jackson}, M. and {Keyes}, C.~D. and {Koekemoer}, A.~M. and {Kotak}, R. and {Le}, D. and {Liska}, D. and {Long}, K.~S. and {Lucey}, J.~R. and {Liu}, M. and {Martin}, N.~F. and {Masci}, G. and {McLean}, B. and {Mindel}, E. and {Misra}, P. and {Morganson}, E. and {Murphy}, D.~N.~A. and {Obaika}, A. and {Narayan}, G. and {Nieto-Santisteban}, M.~A. and {Norberg}, P. and {Peacock}, J.~A. and {Pier}, E.~A. and {Postman}, M. and {Primak}, N. and {Rae}, C. and {Rai}, A. and {Riess}, A. and {Riffeser}, A. and {Rix}, H.~W. and {R{\"o}ser}, S. and {Russel}, R. and {Rutz}, L. and {Schilbach}, E. and {Schultz}, A.~S.~B. and {Scolnic}, D. and {Strolger}, L. and {Szalay}, A. and {Seitz}, S. and {Small}, E. and {Smith}, K.~W. and {Soderblom}, D.~R. and {Taylor}, P. and {Thomson}, R. and {Taylor}, A.~N. and {Thakar}, A.~R. and {Thiel}, J. and {Thilker}, D. and {Unger}, D. and {Urata}, Y. and {Valenti}, J. and {Wagner}, J. and {Walder}, T. and {Walter}, F. and {Watters}, S.~P. and {Werner}, S. and {Wood-Vasey}, W.~M. and {Wyse}, R.},
        title = "{The Pan-STARRS1 Surveys}",
      journal = {arXiv e-prints},
         year = 2016,
        month = dec,
          eid = {arXiv:1612.05560},
        pages = {arXiv:1612.05560},
          doi = {10.48550/arXiv.1612.05560},
archivePrefix = {arXiv},
       eprint = {1612.05560},
 primaryClass = {astro-ph.IM},
       adsurl = {https://ui.adsabs.harvard.edu/abs/2016arXiv161205560C}
}

@ARTICLE{2023A&A...674A...1G,
       author = {{Gaia Collaboration} and {Vallenari}, A. and {Brown}, A.~G.~A. and {Prusti}, T. and {de Bruijne}, J.~H.~J. and {Arenou}, F. and {Babusiaux}, C. and {Biermann}, M. and {Creevey}, O.~L. and {Ducourant}, C. and {Evans}, D.~W. and {Eyer}, L. and {Guerra}, R. and {Hutton}, A. and {Jordi}, C. and {Klioner}, S.~A. and {Lammers}, U.~L. and {Lindegren}, L. and {Luri}, X. and {Mignard}, F. and {Panem}, C. and {Pourbaix}, D. and {Randich}, S. and {Sartoretti}, P. and {Soubiran}, C. and {Tanga}, P. and {Walton}, N.~A. and {Bailer-Jones}, C.~A.~L. and {Bastian}, U. and {Drimmel}, R. and {Jansen}, F. and {Katz}, D. and {Lattanzi}, M.~G. and {van Leeuwen}, F. and {Bakker}, J. and {Cacciari}, C. and {Casta{\~n}eda}, J. and {De Angeli}, F. and {Fabricius}, C. and {Fouesneau}, M. and {Fr{\'e}mat}, Y. and {Galluccio}, L. and {Guerrier}, A. and {Heiter}, U. and {Masana}, E. and {Messineo}, R. and {Mowlavi}, N. and {Nicolas}, C. and {Nienartowicz}, K. and {Pailler}, F. and {Panuzzo}, P. and {Riclet}, F. and {Roux}, W. and {Seabroke}, G.~M. and {Sordo}, R. and {Th{\'e}venin}, F. and {Gracia-Abril}, G. and {Portell}, J. and {Teyssier}, D. and {Altmann}, M. and {Andrae}, R. and {Audard}, M. and {Bellas-Velidis}, I. and {Benson}, K. and {Berthier}, J. and {Blomme}, R. and {Burgess}, P.~W. and {Busonero}, D. and {Busso}, G. and {C{\'a}novas}, H. and {Carry}, B. and {Cellino}, A. and {Cheek}, N. and {Clementini}, G. and {Damerdji}, Y. and {Davidson}, M. and {de Teodoro}, P. and {Nu{\~n}ez Campos}, M. and {Delchambre}, L. and {Dell'Oro}, A. and {Esquej}, P. and {Fern{\'a}ndez-Hern{\'a}ndez}, J. and {Fraile}, E. and {Garabato}, D. and {Garc{\'\i}a-Lario}, P. and {Gosset}, E. and {Haigron}, R. and {Halbwachs}, J. -L. and {Hambly}, N.~C. and {Harrison}, D.~L. and {Hern{\'a}ndez}, J. and {Hestroffer}, D. and {Hodgkin}, S.~T. and {Holl}, B. and {Jan{\ss}en}, K. and {Jevardat de Fombelle}, G. and {Jordan}, S. and {Krone-Martins}, A. and {Lanzafame}, A.~C. and {L{\"o}ffler}, W. and {Marchal}, O. and {Marrese}, P.~M. and {Moitinho}, A. and {Muinonen}, K. and {Osborne}, P. and {Pancino}, E. and {Pauwels}, T. and {Recio-Blanco}, A. and {Reyl{\'e}}, C. and {Riello}, M. and {Rimoldini}, L. and {Roegiers}, T. and {Rybizki}, J. and {Sarro}, L.~M. and {Siopis}, C. and {Smith}, M. and {Sozzetti}, A. and {Utrilla}, E. and {van Leeuwen}, M. and {Abbas}, U. and {{\'A}brah{\'a}m}, P. and {Abreu Aramburu}, A. and {Aerts}, C. and {Aguado}, J.~J. and {Ajaj}, M. and {Aldea-Montero}, F. and {Altavilla}, G. and {{\'A}lvarez}, M.~A. and {Alves}, J. and {Anders}, F. and {Anderson}, R.~I. and {Anglada Varela}, E. and {Antoja}, T. and {Baines}, D. and {Baker}, S.~G. and {Balaguer-N{\'u}{\~n}ez}, L. and {Balbinot}, E. and {Balog}, Z. and {Barache}, C. and {Barbato}, D. and {Barros}, M. and {Barstow}, M.~A. and {Bartolom{\'e}}, S. and {Bassilana}, J. -L. and {Bauchet}, N. and {Becciani}, U. and {Bellazzini}, M. and {Berihuete}, A. and {Bernet}, M. and {Bertone}, S. and {Bianchi}, L. and {Binnenfeld}, A. and {Blanco-Cuaresma}, S. and {Blazere}, A. and {Boch}, T. and {Bombrun}, A. and {Bossini}, D. and {Bouquillon}, S. and {Bragaglia}, A. and {Bramante}, L. and {Breedt}, E. and {Bressan}, A. and {Brouillet}, N. and {Brugaletta}, E. and {Bucciarelli}, B. and {Burlacu}, A. and {Butkevich}, A.~G. and {Buzzi}, R. and {Caffau}, E. and {Cancelliere}, R. and {Cantat-Gaudin}, T. and {Carballo}, R. and {Carlucci}, T. and {Carnerero}, M.~I. and {Carrasco}, J.~M. and {Casamiquela}, L. and {Castellani}, M. and {Castro-Ginard}, A. and {Chaoul}, L. and {Charlot}, P. and {Chemin}, L. and {Chiaramida}, V. and {Chiavassa}, A. and {Chornay}, N. and {Comoretto}, G. and {Contursi}, G. and {Cooper}, W.~J. and {Cornez}, T. and {Cowell}, S. and {Crifo}, F. and {Cropper}, M. and {Crosta}, M. and {Crowley}, C. and {Dafonte}, C. and {Dapergolas}, A. and {David}, M. and {David}, P. and {de Laverny}, P. and {De Luise}, F. and {De March}, R.},
        title = "{Gaia Data Release 3. Summary of the content and survey properties}",
      journal = {\aap},
         year = 2023,
        month = jun,
       volume = {674},
          eid = {A1},
        pages = {A1},
          doi = {10.1051/0004-6361/202243940},
archivePrefix = {arXiv},
       eprint = {2208.00211},
 primaryClass = {astro-ph.GA},
       adsurl = {https://ui.adsabs.harvard.edu/abs/2023A&A...674A...1G}
}

@ARTICLE{2016A&A...595A...1G,
       author = {{Gaia Collaboration} and {Prusti}, T. and {de Bruijne}, J.~H.~J. and {Brown}, A.~G.~A. and {Vallenari}, A. and {Babusiaux}, C. and {Bailer-Jones}, C.~A.~L. and {Bastian}, U. and {Biermann}, M. and {Evans}, D.~W. and {Eyer}, L. and {Jansen}, F. and {Jordi}, C. and {Klioner}, S.~A. and {Lammers}, U. and {Lindegren}, L. and {Luri}, X. and {Mignard}, F. and {Milligan}, D.~J. and {Panem}, C. and {Poinsignon}, V. and {Pourbaix}, D. and {Randich}, S. and {Sarri}, G. and {Sartoretti}, P. and {Siddiqui}, H.~I. and {Soubiran}, C. and {Valette}, V. and {van Leeuwen}, F. and {Walton}, N.~A. and {Aerts}, C. and {Arenou}, F. and {Cropper}, M. and {Drimmel}, R. and {H{\o}g}, E. and {Katz}, D. and {Lattanzi}, M.~G. and {O'Mullane}, W. and {Grebel}, E.~K. and {Holland}, A.~D. and {Huc}, C. and {Passot}, X. and {Bramante}, L. and {Cacciari}, C. and {Casta{\~n}eda}, J. and {Chaoul}, L. and {Cheek}, N. and {De Angeli}, F. and {Fabricius}, C. and {Guerra}, R. and {Hern{\'a}ndez}, J. and {Jean-Antoine-Piccolo}, A. and {Masana}, E. and {Messineo}, R. and {Mowlavi}, N. and {Nienartowicz}, K. and {Ord{\'o}{\~n}ez-Blanco}, D. and {Panuzzo}, P. and {Portell}, J. and {Richards}, P.~J. and {Riello}, M. and {Seabroke}, G.~M. and {Tanga}, P. and {Th{\'e}venin}, F. and {Torra}, J. and {Els}, S.~G. and {Gracia-Abril}, G. and {Comoretto}, G. and {Garcia-Reinaldos}, M. and {Lock}, T. and {Mercier}, E. and {Altmann}, M. and {Andrae}, R. and {Astraatmadja}, T.~L. and {Bellas-Velidis}, I. and {Benson}, K. and {Berthier}, J. and {Blomme}, R. and {Busso}, G. and {Carry}, B. and {Cellino}, A. and {Clementini}, G. and {Cowell}, S. and {Creevey}, O. and {Cuypers}, J. and {Davidson}, M. and {De Ridder}, J. and {de Torres}, A. and {Delchambre}, L. and {Dell'Oro}, A. and {Ducourant}, C. and {Fr{\'e}mat}, Y. and {Garc{\'\i}a-Torres}, M. and {Gosset}, E. and {Halbwachs}, J. -L. and {Hambly}, N.~C. and {Harrison}, D.~L. and {Hauser}, M. and {Hestroffer}, D. and {Hodgkin}, S.~T. and {Huckle}, H.~E. and {Hutton}, A. and {Jasniewicz}, G. and {Jordan}, S. and {Kontizas}, M. and {Korn}, A.~J. and {Lanzafame}, A.~C. and {Manteiga}, M. and {Moitinho}, A. and {Muinonen}, K. and {Osinde}, J. and {Pancino}, E. and {Pauwels}, T. and {Petit}, J. -M. and {Recio-Blanco}, A. and {Robin}, A.~C. and {Sarro}, L.~M. and {Siopis}, C. and {Smith}, M. and {Smith}, K.~W. and {Sozzetti}, A. and {Thuillot}, W. and {van Reeven}, W. and {Viala}, Y. and {Abbas}, U. and {Abreu Aramburu}, A. and {Accart}, S. and {Aguado}, J.~J. and {Allan}, P.~M. and {Allasia}, W. and {Altavilla}, G. and {{\'A}lvarez}, M.~A. and {Alves}, J. and {Anderson}, R.~I. and {Andrei}, A.~H. and {Anglada Varela}, E. and {Antiche}, E. and {Antoja}, T. and {Ant{\'o}n}, S. and {Arcay}, B. and {Atzei}, A. and {Ayache}, L. and {Bach}, N. and {Baker}, S.~G. and {Balaguer-N{\'u}{\~n}ez}, L. and {Barache}, C. and {Barata}, C. and {Barbier}, A. and {Barblan}, F. and {Baroni}, M. and {Barrado y Navascu{\'e}s}, D. and {Barros}, M. and {Barstow}, M.~A. and {Becciani}, U. and {Bellazzini}, M. and {Bellei}, G. and {Bello Garc{\'\i}a}, A. and {Belokurov}, V. and {Bendjoya}, P. and {Berihuete}, A. and {Bianchi}, L. and {Bienaym{\'e}}, O. and {Billebaud}, F. and {Blagorodnova}, N. and {Blanco-Cuaresma}, S. and {Boch}, T. and {Bombrun}, A. and {Borrachero}, R. and {Bouquillon}, S. and {Bourda}, G. and {Bouy}, H. and {Bragaglia}, A. and {Breddels}, M.~A. and {Brouillet}, N. and {Br{\"u}semeister}, T. and {Bucciarelli}, B. and {Budnik}, F. and {Burgess}, P. and {Burgon}, R. and {Burlacu}, A. and {Busonero}, D. and {Buzzi}, R. and {Caffau}, E. and {Cambras}, J. and {Campbell}, H. and {Cancelliere}, R. and {Cantat-Gaudin}, T. and {Carlucci}, T. and {Carrasco}, J.~M. and {Castellani}, M. and {Charlot}, P. and {Charnas}, J. and {Charvet}, P. and {Chassat}, F. and {Chiavassa}, A. and {Clotet}, M. and {Cocozza}, G. and {Collins}, R.~S. and {Collins}, P. and {Costigan}, G.},
        title = "{The Gaia mission}",
      journal = {\aap},
         year = 2016,
        month = nov,
       volume = {595},
          eid = {A1},
        pages = {A1},
          doi = {10.1051/0004-6361/201629272},
archivePrefix = {arXiv},
       eprint = {1609.04153},
 primaryClass = {astro-ph.IM},
       adsurl = {https://ui.adsabs.harvard.edu/abs/2016A&A...595A...1G}
}

@ARTICLE{2010AJ....139.1782L,
       author = {{Lang}, Dustin and {Hogg}, David W. and {Mierle}, Keir and {Blanton}, Michael and {Roweis}, Sam},
        title = "{Astrometry.net: Blind Astrometric Calibration of Arbitrary Astronomical Images}",
      journal = {\aj},
         year = 2010,
        month = may,
       volume = {139},
       number = {5},
        pages = {1782-1800},
          doi = {10.1088/0004-6256/139/5/1782},
archivePrefix = {arXiv},
       eprint = {0910.2233},
 primaryClass = {astro-ph.IM},
       adsurl = {https://ui.adsabs.harvard.edu/abs/2010AJ....139.1782L}
}

@ARTICLE{2022JApA...43...10K,
       author = {{Kumar}, Brajesh and {Negi}, Vibhore and {Ailawadhi}, Bhavya and {Mishra}, Sapna and {Pradhan}, Bikram and {Misra}, Kuntal and {Hickson}, Paul and {Surdej}, Jean},
        title = "{Upcoming 4m ILMT facility and data reduction pipeline testing}",
      journal = {Journal of Astrophysics and Astronomy},
         year = 2022,
        month = jun,
       volume = {43},
       number = {1},
          eid = {10},
        pages = {10},
          doi = {10.1007/s12036-021-09795-3},
archivePrefix = {arXiv},
       eprint = {2112.01209},
 primaryClass = {astro-ph.IM},
       adsurl = {https://ui.adsabs.harvard.edu/abs/2022JApA...43...10K}
}

@INPROCEEDINGS{1980SPIE..264...20M,
       author = {{McGraw}, J.~T. and {Angel}, J.~R.~P. and {Sargent}, T.~A.},
        title = "{A charge-coupled device /CCD/ transit-telescope survey for galactic and extragalactic variability and polarization}",
    booktitle = {Conference on Applications of Digital Image Processing to Astronomy},
         year = 1980,
       editor = {{Elliott}, D.~A.},
       series = {Society of Photo-Optical Instrumentation Engineers (SPIE) Conference Series},
       volume = {264},
        month = jan,
        pages = {20-28},
          doi = {10.1117/12.959777},
       adsurl = {https://ui.adsabs.harvard.edu/abs/1980SPIE..264...20M}
}

@ARTICLE{1984MNRAS.210..979H,
       author = {{Hall}, P. and {Mackay}, C.~D.},
        title = "{Faint galaxy number - magnitude counts at high galactic latitude -I.}",
      journal = {MNRAS},
         year = 1984,
        month = oct,
       volume = {210},
        pages = {979-992},
          doi = {10.1093/mnras/210.4.979},
       adsurl = {https://ui.adsabs.harvard.edu/abs/1984MNRAS.210..979H}
}

@ARTICLE{1998PASP..110.1081H,
       author = {{Hickson}, Paul and {Richardson}, E. Harvey},
        title = "{A Curvature-compensated Corrector for Drift-Scan Observations}",
      journal = {PASP},
         year = 1998,
        month = sep,
       volume = {110},
       number = {751},
        pages = {1081-1086},
          doi = {10.1086/316230},
archivePrefix = {arXiv},
       eprint = {astro-ph/9806303},
 primaryClass = {astro-ph},
       adsurl = {https://ui.adsabs.harvard.edu/abs/1998PASP..110.1081H}
}

@ARTICLE{fits_wcs_1,
       author = {{Greisen}, E.~W. and {Calabretta}, M.~R.},
        title = "{Representations of world coordinates in FITS}",
      journal = {\aap},
         year = 2002,
        month = dec,
       volume = {395},
        pages = {1061-1075},
          doi = {10.1051/0004-6361:20021326},
archivePrefix = {arXiv},
       eprint = {astro-ph/0207407},
 primaryClass = {astro-ph},
       adsurl = {https://ui.adsabs.harvard.edu/abs/2002A&A...395.1061G}
}

@ARTICLE{fits_wcs_2,
       author = {{Calabretta}, M.~R. and {Greisen}, E.~W.},
        title = "{Representations of celestial coordinates in FITS}",
      journal = {\aap},
         year = 2002,
        month = dec,
       volume = {395},
        pages = {1077-1122},
          doi = {10.1051/0004-6361:20021327},
archivePrefix = {arXiv},
       eprint = {astro-ph/0207413},
 primaryClass = {astro-ph},
       adsurl = {https://ui.adsabs.harvard.edu/abs/2002A&A...395.1077C}
}

@INPROCEEDINGS{fits_sip_convention,
       author = {{Shupe}, D.~L. and {Moshir}, Mehdrdad and {Li}, J. and {Makovoz}, D. and {Narron}, R. and {Hook}, R.~N.},
        title = "{The SIP Convention for Representing Distortion in FITS Image Headers}",
    booktitle = {Astronomical Data Analysis Software and Systems XIV},
         year = 2005,
       editor = {{Shopbell}, P. and {Britton}, M. and {Ebert}, R.},
       series = {Astronomical Society of the Pacific Conference Series},
       volume = {347},
        month = dec,
        pages = {491},
       adsurl = {https://ui.adsabs.harvard.edu/abs/2005ASPC..347..491S}
}

@INPROCEEDINGS{2019LPICo2109.6066H,
       author = {{Hickson}, P.},
        title = "{OCS: A Flexible Observatory Control System for Robotic Telescopes with Application to Detection and Characterization of Orbital Debris}",
    booktitle = {First International Orbital Debris Conference},
         year = 2019,
       series = {LPI Contributions},
       volume = {2109},
        month = dec,
          eid = {6066},
        pages = {6066},
       adsurl = {https://ui.adsabs.harvard.edu/abs/2019LPICo2109.6066H}
}

@ARTICLE{2013A&A...558A..33A,
       author = {{Astropy Collaboration} and {Robitaille}, Thomas P. and {Tollerud}, Erik J. and {Greenfield}, Perry and {Droettboom}, Michael and {Bray}, Erik and {Aldcroft}, Tom and {Davis}, Matt and {Ginsburg}, Adam and {Price-Whelan}, Adrian M. and {Kerzendorf}, Wolfgang E. and {Conley}, Alexander and {Crighton}, Neil and {Barbary}, Kyle and {Muna}, Demitri and {Ferguson}, Henry and {Grollier}, Fr{\'e}d{\'e}ric and {Parikh}, Madhura M. and {Nair}, Prasanth H. and {Unther}, Hans M. and {Deil}, Christoph and {Woillez}, Julien and {Conseil}, Simon and {Kramer}, Roban and {Turner}, James E.~H. and {Singer}, Leo and {Fox}, Ryan and {Weaver}, Benjamin A. and {Zabalza}, Victor and {Edwards}, Zachary I. and {Azalee Bostroem}, K. and {Burke}, D.~J. and {Casey}, Andrew R. and {Crawford}, Steven M. and {Dencheva}, Nadia and {Ely}, Justin and {Jenness}, Tim and {Labrie}, Kathleen and {Lim}, Pey Lian and {Pierfederici}, Francesco and {Pontzen}, Andrew and {Ptak}, Andy and {Refsdal}, Brian and {Servillat}, Mathieu and {Streicher}, Ole},
        title = "{Astropy: A community Python package for astronomy}",
      journal = {\aap},
         year = 2013,
        month = oct,
       volume = {558},
          eid = {A33},
        pages = {A33},
          doi = {10.1051/0004-6361/201322068},
archivePrefix = {arXiv},
       eprint = {1307.6212},
 primaryClass = {astro-ph.IM},
       adsurl = {https://ui.adsabs.harvard.edu/abs/2013A&A...558A..33A}
}

@ARTICLE{2002A&A...388..712V,
       author = {{Vangeyte}, B. and {Manfroid}, J. and {Surdej}, J.},
        title = "{Study of CCD mosaic configurations for the ILMT: Astrometry and photometry of point sources in the absence of a TDI corrector}",
      journal = {\aap},
         year = 2002,
        month = jun,
       volume = {388},
        pages = {712-731},
          doi = {10.1051/0004-6361:20020516},
       adsurl = {https://ui.adsabs.harvard.edu/abs/2002A&A...388..712V}
}

@software{2016ascl.soft06014N,
       author = {{Newville}, Matthew and {Stensitzki}, Till and {Allen}, Daniel B. and {Rawlik}, Michal and {Ingargiola}, Antonino and {Nelson}, Andrew},
        title = "{Lmfit: Non-Linear Least-Square Minimization and Curve-Fitting for Python}",
 howpublished = {Astrophysics Source Code Library, record ascl:1606.014},
         year = 2016,
        month = jun,
          eid = {ascl:1606.014},
archivePrefix = {ascl},
       eprint = {1606.014},
       adsurl = {https://ui.adsabs.harvard.edu/abs/2016ascl.soft06014N}
}
\bibliographystyle{aasjournalv7}



\end{document}